# Balanced Datasets for IoT IDS

Alaa Alhowaide*
Computer Science Department
Memorial University of Newfoundland
St. John's, NL, Canada
azalhowaide@mun.ca

Izzat Alsmadi
Department of Computing and Cybersecurity
Texas A&M University-San Antonio
San Antonio, Texas, United States
ialsmadi@tamusa.edu

Jian Tang
Computer Science Department
Memorial University of Newfoundland
St. John's, NL, Canada
jian@mun.ca

**Abstract**

As the Internet of Things (IoT) continues to grow, cyberattacks are becoming increasingly common. The security of IoT networks relies heavily on intrusion detection systems (IDSs). The development of an IDS that is accurate and efficient is a challenging task. As a result, this challenge is made more challenging by the absence of balanced datasets for training and testing the proposed IDS. In this study, four commonly used datasets are visualized and analyzed visually. Moreover, it proposes a sampling algorithm that generates a sample that represents the original dataset. In addition, it proposes an algorithm to generate a balanced dataset. Researchers can use this paper as a starting point when investigating cybersecurity and machine learning.

The proposed sampling algorithms showed reliability in generating well-representing and balanced samples from NSL-KDD, UNSW-NB15, BotNetIoT-01, and BoTIoT datasets.

## 1   Introduction

A rapid growth of applications in our lives depending on Internet of Things (IoT). IoT applications connect everything from thermostats, toaster, fridge to complete systems to the Internet [1]. Most of the common IoT applications include personal healthcare [2], smart transportations [3] [2], smart grid [4], smart industrial automation [5], and intelligent emergency response systems [2] ranging from monitoring to decision making. All these systems aim to provide a more comfortable lifestyle and improve our capabilities to experience a considerably better life [3], which looks "smart.". The number of connected devices reached 9.5 billion devices at the end of 2019, according to IoT Analytics [6]. IoT Analytics expects the previous figure to reach 22 billion by 2025 [7]. This rapid growth aligned with lake of standers and low-cost manufacturing enormous increased the number and types of security threats [8]. Consequently, IoT network security becomes mandatory.

IDS is the first defense line against cyber-attacks. It refers to a device or software strategically allocated at a specific point on a network to monitor all the traffic [9]. IDS reports malicious behaviors to the network administrator, stop the malicious behaviors, and encompass intruders' identification [8]. IDS are categorized into an anomaly, and signature detection, and hybrid. The anomaly-based detection depends on the behavioral methods in which it defines two types of behaviors; the normal and abnormal behaviors [10]. Most of IDS types of highly depends on Machine Learning (ML) [9], [11], and [12], . ML plays a significant role in developing IDSs to detect malicious threats. The performance of ML models highly depends on the dataset(s) that are used to train, validate, and test the MLs. With the absence of balanced datasets and lake of standards and details of how samples are selected from the training/testing datasets, results become unreliable.

The contributions of this research are (i) visual analysis of commonly used datasets; (ii) sampling algorithm that generates a representative sample of the original datasets (iii) sampling algorithm that generates representative balanced samples from the imbalanced datasets; (iii) two methods to measure the statistical significance of the produced results.

Section two provides a literature review. Section three explains the proposed sampling algorithms, while sections four and five discuss experiments and results, respectively. Lastly, section six concludes the study.

## 2  Datasets in literature

This research considers four datasets: NSL-KDD [13], UNSW-NB15 [14], BotNetIoT-01 [15], and BoTIoT [16] datasets. The NSL-KDD dataset is the most commonly used in intrusion detection, which contains regular network traffic. Similarly, UNSW-NB15 contains regular network traffic. NSL-KDD and UNSW-NB15 represent a benchmark for intrusion detection. In this regard, they allow the current research to be compared with previous studies. BotNetIoT-01 dataset contains the traffic of nine IoT devices. Mirai and Gafgyt botnet attacks are the only attacks in the BotNetIoT-01 dataset. BotNetIoT-01 dataset has 23 feature. This dataset is available at [17]. BoTIoT is a recently published dataset of simulated IoT network traffic. BoTIoT has a variety of recent/new IoT attacks. A detailed analysis of these datasets features and traffic is available at [15] and [9].

## 3  Proposed sampling methods

This section presents two sampling algorithms. Algorithm 1 select a random sample that is similar to the original dataset. It uses the permutation to increase randomness of selecting the sample. Then it simply selects the first *num* of records in the reindexed dataset. After selecting a random sample, it checks the similarity of the selected labels' distributions with the original dataset labels' distribution. If they were similar, it returns the random sample.

**Algorithm 1:** The algorithm for sampling
**Input:** *DS:* dataset, *num*: sample size
**Output:** *S*: Sample of size *num*
**Procedure: getSample**(*DS, num*)
1  $S \leftarrow \emptyset$
2  *Repeat as S is not similar to DS*
3     $S = DS$.reindex(np.random.permutation(dataframe.index))
4     $S = S$.head(*num*)
5  Return $S$

Algorithm 2 generates a balanced sample by first identifying the least represented class label in the original dataset, referred as *MinLabel*. Then, it selects from the dominant class label a number or records equals to the number of *MinLabel* using Algorithm 1.

**Algorithm 2:** The algorithm for generating a balanced sample
**Input:** *DS:* dataset
**Output:** *B* : Balanced sample

**Procedure: getBalancedSample**(*DS*)
1    $B \leftarrow \emptyset$
2    Define minimum class label, *MinLabel*
3    *minClassSet*=DS['label']== *MinLabel*]
4    *S*=**getSample**(DS['label']!= *MinLabel*], minimum class label count)
5    *B=concat(minClassSet, S)*
6    Return *B*

## 4  Experiments

All experiments were implemented using Python 3.8.8. A label feature was added or edited in every dataset to represent the normal and attack traffics by 0 and 1, respectively. Furthermore, all the categorical features values were transformed into numerical values to facilitate the visual representation using the PCA. The duplicate records were removed from all the datasets as well. In NSL-KDD, the features num_outbound_cmds and is_hot_login were dropped because they have zero value for all records. The generated balanced samples are available at [**].

The Z-test experiments were executed with a freedom value equal to zero. The alternative hypothesis (H1) was: The difference in means is not equal to zero. The experiments considered two methods to measure the significant difference between the datasets and their balanced samples. The first method considered the original features of a corresponding dataset. The method measured the similarity using Z-test for each feature. If all features were found similar, then the balances sample is considered as a good representative sample of the original dataset.

The second method considered three dimensional PCA model to represent each of the original datasets and the balanced sample. Similarly, it tests the similarity of each of the PCAs' dimensions. The two sets are similar if all the PCA's dimensions are similar.

## 5  Results

Figures and Tables from 1 to 12 display the distributions and counts, respectively, for each traffic type in the datasets and the generated samples. The blue color in the pie charts stands for normal traffic, while other colors stand for different attacks. The normal traffic is the dominant class in NSL-KDD and UNSW-NB15 datasets, while it is the opposite case for BoTIoT BotNetIoT-01 datasets.

Significantly, the 50% sample pie charts are identical to the original datasets for all the datasets. This similarity of distribution leads to the success of Algorithm 1 proposed in section 3. Notably, the attacks distribution of BotNetIoT-01 and BoTIoT balanced samples is quite similar to the corresponding original datasets. It is notable that the highly represented attack type in the original dataset still reserves its representation in the 50% and balanced datasets.

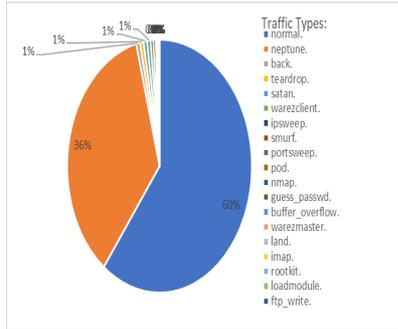
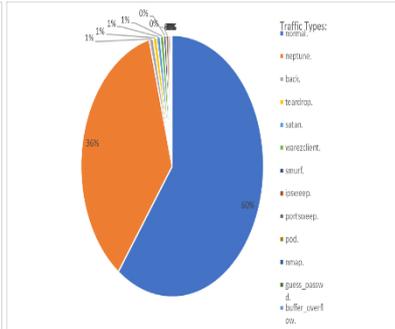
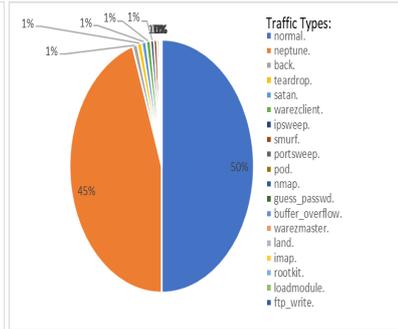

Fig. 1 NSL full dataset traffic distribution.

Fig. 2 NSL 50% dataset traffic distribution.

Fig. 3 NSL balanced dataset traffic distribution.

Table 1 NSL full dataset traffic counts.

| Traffic Type | count | % |
| --- | --- | --- |
| Normal. | 87832 | 0.60329977 |
| neptune. | 51820 | 0.35594082 |
| back. | 968 | 0.00664899 |
| teardrop. | 918 | 0.00630555 |
| satan. | 906 | 0.00622313 |
| warezclient. | 893 | 0.00613383 |
| ipsweep. | 651 | 0.00447158 |
| smurf. | 641 | 0.0044029 |
| portsweep. | 416 | 0.00285742 |
| pod. | 206 | 0.00141497 |
| nmap. | 158 | 0.00108527 |
| guess_passwd. | 53 | 0.00036405 |
| buffer_overflow | 30 | 0.00020606 |
| warezmaster. | 20 | 0.00013738 |
| land. | 19 | 0.00013051 |
| imap. | 12 | 8.2426E-05 |
| rootkit. | 10 | 6.8688E-05 |
| loadmodule. | 9 | 6.1819E-05 |
| ftp_write. | 8 | 5.495E-05 |
| multihop. | 7 | 4.8082E-05 |
| phf. | 4 | 2.7475E-05 |
| perl. | 3 | 2.0606E-05 |
| spy. | 2 | 1.3738E-05 |
| Total | 145586 | |

Table 2 NSL 50% dataset traffic counts.

| Traffic Type | count | % |
| --- | --- | --- |
| Normal. | 43717 | 0.60056599 |
| neptune. | 26118 | 0.35879824 |
| back. | 506 | 0.00695122 |
| teardrop. | 464 | 0.00637424 |
| satan. | 461 | 0.00633303 |
| warezclient. | 451 | 0.00619565 |
| smurf. | 304 | 0.00417623 |
| ipsweep. | 302 | 0.00414875 |
| portsweep. | 213 | 0.00292611 |
| pod. | 104 | 0.00142871 |
| nmap. | 79 | 0.00108527 |
| guess_passwd. | 21 | 0.00028849 |
| buffer_overflow | 16 | 0.0002198 |
| warezmaster. | 11 | 0.00015111 |
| rootkit. | 6 | 8.2426E-05 |
| imap. | 5 | 6.8688E-05 |
| ftp_write. | 4 | 5.495E-05 |
| land. | 4 | 5.495E-05 |
| multihop. | 4 | 5.495E-05 |
| phf. | 1 | 1.3738E-05 |
| loadmodule. | 1 | 1.3738E-05 |
| perl. | 1 | 1.3738E-05 |
| Total | 72793 | |

Table 3 NSL balanced dataset traffic counts.

| Traffic Type | count | % |
| --- | --- | --- |
| Normal. | 57754 | 0.5 |
| neptune. | 51820 | 0.44862693 |
| back. | 968 | 0.00838037 |
| teardrop. | 918 | 0.0079475 |
| satan. | 906 | 0.00784361 |
| warezclient. | 893 | 0.00773107 |
| ipsweep. | 651 | 0.00563597 |
| smurf. | 641 | 0.0055494 |
| portsweep. | 416 | 0.00360148 |
| pod. | 206 | 0.00178343 |
| nmap. | 158 | 0.00136787 |
| guess_passwd. | 53 | 0.00045884 |
| buffer_overflow | 30 | 0.00025972 |
| warezmaster. | 20 | 0.00017315 |
| land. | 19 | 0.00016449 |
| imap. | 12 | 0.00010389 |
| rootkit. | 10 | 8.6574E-05 |
| loadmodule. | 9 | 7.7917E-05 |
| ftp_write. | 8 | 6.9259E-05 |
| multihop. | 7 | 6.0602E-05 |
| phf. | 4 | 3.463E-05 |
| perl. | 3 | 2.5972E-05 |
| spy. | 2 | 1.7315E-05 |
| Total | 115508 | |

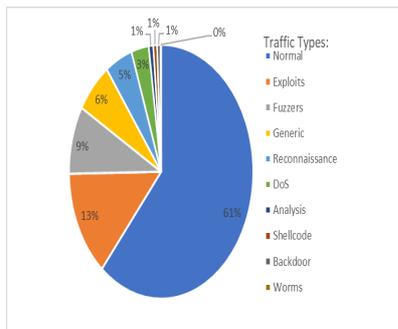
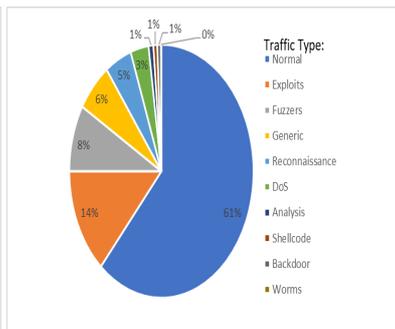
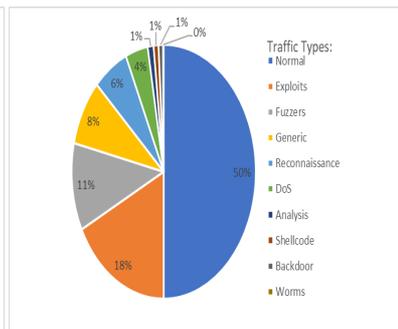

Fig. 4 UNSW-NB15 full dataset traffic distribution.

Fig. 5 UNSW-NB15 50% dataset traffic distribution.

Fig. 6 UNSW-NB15 balanced dataset traffic distribution.

Table 4 UNSW-NB15 full dataset traffic counts.

| Traffic Type | count | % |
| --- | --- | --- |
| Normal | 34205 | 0.61141499 |
| Exploits | 7609 | 0.13601101 |
| Fuzzers | 4838 | 0.08647934 |
| Generic | 3657 | 0.06536894 |
| Reconnaissance | 2703 | 0.04831617 |
| DoS | 1718 | 0.03070928 |
| Analysis | 446 | 0.00797226 |
| Shellcode | 378 | 0.00675676 |
| Backdoor | 346 | 0.00618476 |
| Worms | 44 | 0.0007865 |
| Total | 55944 | |

Table 5 UNSW-NB15 50% dataset traffic counts.

| Traffic Type | count | % |
| --- | --- | --- |
| Normal | 17185 | 0.61436436 |
| Exploits | 3797 | 0.13574289 |
| Fuzzers | 2380 | 0.08508509 |
| Generic | 1769 | 0.06324181 |
| Reconnaissance | 1321 | 0.0472258 |
| DoS | 892 | 0.03188903 |
| Analysis | 236 | 0.00843701 |
| Shellcode | 192 | 0.00686401 |
| Backdoor | 183 | 0.00654226 |
| Worms | 17 | 0.00060775 |
| Total | 27972 | |

Table 6 UNSW-NB15 balanced dataset traffic counts.

| Traffic Type | count | % |
| --- | --- | --- |
| Normal | 21739 | 0.5 |
| Exploits | 7609 | 0.17500805 |
| Fuzzers | 4838 | 0.11127467 |
| Generic | 3657 | 0.0841115 |
| Reconnaissance | 2703 | 0.06216937 |
| DoS | 1718 | 0.03951424 |
| Analysis | 446 | 0.01025806 |
| Shellcode | 378 | 0.00869405 |
| Backdoor | 346 | 0.00795805 |
| Worms | 44 | 0.00101201 |
| Total | 43478 | |

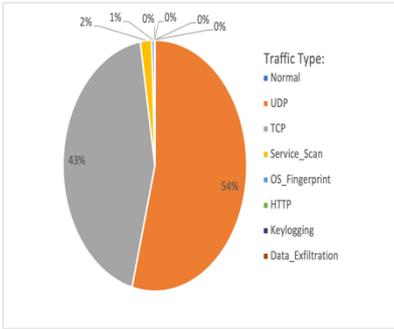
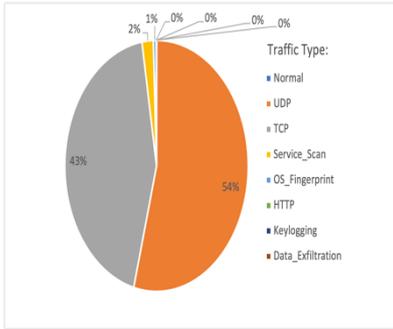
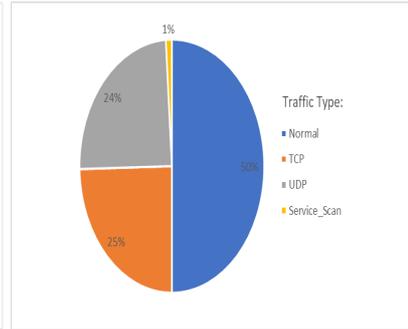

Fig. 7 BotIoT full dataset traffic distribution.

Fig. 8 BotIoT 50% dataset traffic distribution.

Fig. 9 BotIoT balanced dataset traffic distribution.

Table 7 BotIoT full dataset traffic counts.

Table 8 BotIoT 50% dataset traffic counts.

Table 9 BotIoT balanced dataset traffic counts.

| Traffic Type | count | % |
| --- | --- | --- |
| Normal | 477 | 0.00013003 |
| UDP | 1981230 | 0.54006218 |
| TCP | 1593180 | 0.43428389 |
| Service_Scan | 73168 | 0.01994482 |
| OS_Fingerprint | 17914 | 0.00488317 |
| HTTP | 2474 | 0.00067439 |
| Keylogging | 73 | 1.9899E-05 |
| Data_Exfiltration | 6 | 1.6355E-06 |
| Total | 3668522 | |

| Traffic Type | count | % |
| --- | --- | --- |
| Normal | 231 | 0.000125936 |
| UDP | 990566 | 0.540035469 |
| TCP | 796673 | 0.434329139 |
| Service_Scan | 36604 | 0.019955721 |
| OS_Fingerprint | 8916 | 0.004860813 |
| HTTP | 1239 | 0.000675476 |
| Keylogging | 29 | 1.58102E-05 |
| Data_Exfiltration | 3 | 1.63554E-06 |
| Total | 1834261 | |

| Traffic Type | count | % |
| --- | --- | --- |
| Normal | 477 | 0.5 |
| TCP | 235 | 0.24633124 |
| UDP | 232 | 0.24318658 |
| Service_Scan | 10 | 0.01048218 |
| Total | 954 | |

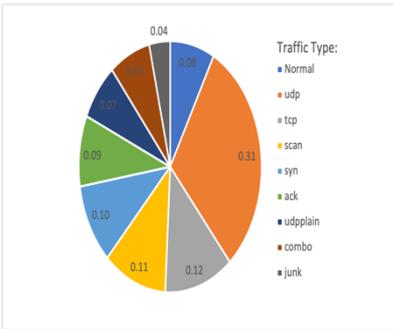
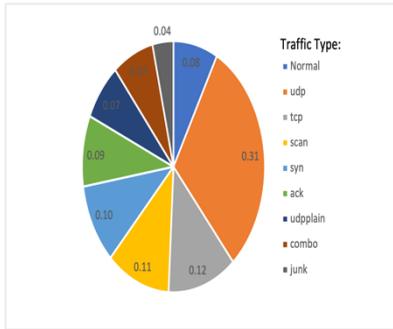
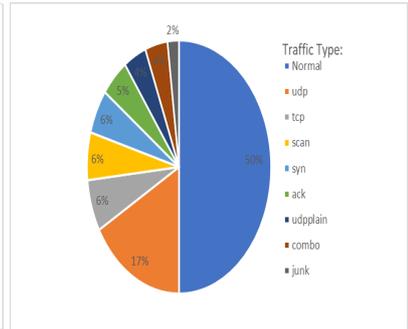

Fig. 10 BotNetIoT-01 full dataset traffic distribution.

Fig. 11 BotNetIoT-01 50% dataset traffic distribution.

Fig. 12 BotNetIoT-01 balanced dataset traffic distribution.

Table 10 BotNetIoT-01 full dataset traffic counts.

Table 11 BotNetIoT-01 50% dataset traffic counts.

Table 12 BotNetIoT-01 balanced dataset traffic counts.

| Traffic Type | count | % |
| --- | --- | --- |
| Normal | 555932 | 0.07871485 |
| udp | 2176365 | 0.30815325 |
| tcp | 859850 | 0.12174685 |
| scan | 793090 | 0.11229424 |
| syn | 733299 | 0.10382839 |
| ack | 643821 | 0.09115913 |
| udpplain | 523304 | 0.07409503 |
| combo | 515156 | 0.07294135 |
| junk | 261789 | 0.03706691 |
| Total | 7062606 | |

| Traffic Type | count | % |
| --- | --- | --- |
| Normal | 277631 | 0.07861999 |
| udp | 1088615 | 0.30827573 |
| tcp | 429603 | 0.12165566 |
| scan | 397213 | 0.11248341 |
| syn | 366084 | 0.10366825 |
| ack | 322252 | 0.09125583 |
| udpplain | 261385 | 0.07401942 |
| combo | 257771 | 0.072996 |
| junk | 130749 | 0.03702571 |
| Total | 3531303 | |

| Traffic Type | count | % |
| --- | --- | --- |
| Normal | 555932 | 0.5 |
| udp | 186062 | 0.16734241 |
| tcp | 73405 | 0.06601977 |
| scan | 67626 | 0.06082219 |
| syn | 62917 | 0.05658696 |
| ack | 54864 | 0.04934416 |
| udpplain | 44624 | 0.04013441 |
| combo | 44091 | 0.03965503 |
| junk | 22343 | 0.02009508 |
| Total | 1111864 | |

To increase the reliability of the produced results by algorithms 1 and 2, we executed Z-test to measure the statistical significance. Z-test was applied in two ways to measure if there is a significant difference between the original dataset and the balanced dataset. Table 13 shows the two methods results of comparing the 50% and the balanced samples compared with their corresponding original datasets. Only the NSL 50% sample was found similar using the first method. On the other hand, all samples were found similar using the second method.

Table 13 Z-test results using All-feature and PCA comparisons.

| Dataset | Sample | Features | PCA |
|---|---|---|---|
| NSL | 50% | similar | similar |
| | Balanced DS | 13 similar, 27 different features | similar |
| NB15 | 50% DS | 6 similar, 37 different | similar |
| | Balanced DS | 4 similar, 39 different features | similar |
| Botnetiot | 50% DS | 24 similar, 0 different | similar |
| | Balanced DS | 0 similar, 24 different features | similar |
| Botiot | 50% DS | 43 same, 0 different | similar |
| | Balanced DS | 4 similar, 39 different features | similar |

To better understand these results, we plot a visual representation of the original datasets, 50%, and balanced samples using the PCA. Table 14 presents all the 3D PCA plot for all datasets and samples. Also, the explained variance of each dimension of the PCA and the accumulative explained variance are displayed above each plot. Interestingly, the samples have similar plot as their corresponding datasets, without exception. From Table 14, the second method that used PCA seems to be more accurate.

Table 14 3D visualization of all datasets and samples using PCA.

| Dataset | Full | 50% sample | Balanced sample |
|---|---|---|---|
| NSL | Dim Var=[0.99888489 0.99999954 0.99999999]<br>Acc Var=0.9999999897744162 | Dim Var=[0.99945449 0.99999977 0.99999999]<br>Acc Var=0.999999994890656 | Dim Var=[0.99894055 0.99999965 0.99999999]<br>Acc Var0.9999999918070609 |
| UNSW-NB15 | Dim Var= [0.74721778 0.99154115 0.99999804]<br>Acc Var=0.9999980361116154 | Dim Var= [0.74798519 0.99129133 0.99999805]<br>Acc Var=0.9999980485289511 | Dim Var= [0.74993195 0.9919089 0.99999836]<br>Acc Var=0.9999983606509643 |
| BotIoT | Dim Var=[0.61936997 0.87203041 0.97291457]<br>Acc Var=0.9729145691941684 | Dim Var= [0.57665246 0.83934756 0.96824873]<br>Acc Var=0.9682487261636522 | Dim Var= [0.83554185 0.97786429 0.99720171]<br>Acc Var= 0.9972017146909109 |
| BotNetIoT-01 | Dim Var=[1. 1. 1.]<br>Acc Var=1 | Dim Var=[1. 1. 1.]<br>Acc Var=0.9999999999996793 | Dim Var=[1. 1. 1.]<br>Acc Var=0.9999999999998491 |

# 6 Conclusion

As IoT network evolves and expands, the need for reliable and efficient IDSs increases. Therefore, it is mandatory to have high-quality datasets to train and test these security systems. It is significant to have balanced datasets to generate reliable results. In this research, four commonly used datasets to build IDSs were used to extract balanced samples. Two novel sampling algorithms were proposed. Furthermore, two methods that measure the goodness of the generated samples were proposed.

Training the IDS on balanced datasets has a significant role in learning reliable models. It increases our confidence in the detection model's results. The proposed algorithms showed the ability to generate good representing samples out from the used datasets based on the Z-test and the 3D PCA plots.

# 7 References

[1] S. Chawla and G. Thamilarasu, "Security as a service: real-time intrusion detection in internet of things," in *Proceedings of the Fifth Cybersecurity Symposium on - CyberSec '18*, Coeur d' Alene, Idaho, Apr. 2018, pp. 1–4. doi: 10.1145/3212687.3212872.

[2] L. Atzori, A. Lera, and G. Morabito, "The Internet of Things: A survey | Elsevier Enhanced Reader," *Computer networks*, vol. 54, no. 15, pp. 2787–2805, 2010, doi: 10.1016/j.comnet.2010.05.010.

[3] A. Al-Fuqaha, M. Guizani, M. Mohammadi, M. Aledhari, and M. Ayyash, "Internet of Things: A Survey on Enabling Technologies, Protocols, and Applications," *IEEE Communications Surveys Tutorials*, vol. 17, no. 4, pp. 2347–2376, Fourthquarter 2015, doi: 10.1109/COMST.2015.2444095.

[4] K. J. Kaur and A. Hahn, "Exploring ensemble classifiers for detecting attacks in the smart grids," in *Proceedings of the Fifth Cybersecurity Symposium*, Coeur d' Alene Idaho, Apr. 2018, pp. 1–4. doi: 10.1145/3212687.3212873.

[5] S. Shen, L. Huang, H. Zhou, S. Yu, E. Fan, and Q. Cao, "Multistage Signaling Game-Based Optimal Detection Strategies for Suppressing Malware Diffusion in Fog-Cloud-Based IoT Networks," *IEEE Internet of Things Journal*, vol. 5, no. 2, pp. 1043–1054, Apr. 2018, doi: 10.1109/JIOT.2018.2795549.

[6] K. Lueth, "IoT 2019 in Review: The 10 Most Relevant IoT Developments of the Year." https://iot-analytics.com/iot-2019-in-review/ (accessed May 27, 2020).

[7] K. Lueth, "State of the IoT 2018: Number of IoT devices now at 7B – Market accelerating." https://iot-analytics.com/state-of-the-iot-update-q1-q2-2018-number-of-iot-devices-now-7b/ (accessed May 27, 2020).

[8] A. Alhowaide, I. Alsmadi, and J. Tang, "Ensemble Detection Model for IoT IDS," *Internet of Things*, p. 100435, Jul. 2021, doi: 10.1016/j.iot.2021.100435.

[9] A. Alhowaide, I. Alsmadi, and J. Tang, "Towards the design of real-time autonomous IoT NIDS," *Cluster Comput*, Jan. 2021, doi: 10.1007/s10586-021-03231-5.

[10] M. Aldwairi, W. Mardini, and A. Alhowaide, "Anomaly Payload Signature Generation System Based on Efficient Tokenization Methodology," *International Journal on Communications Antenna and Propagation (IRECAP) (2018)*, Nov. 2018.

[11] A. Alhowaide, I. Alsmadi, and J. Tang, "PCA, Random-Forest and Pearson Correlation for Dimensionality Reduction in IoT IDS," in *2020 IEEE International IOT, Electronics and Mechatronics Conference (IEMTRONICS)*, Sep. 2020, pp. 1–6. doi: 10.1109/IEMTRONICS51293.2020.9216388.

[12] A. Alhowaide, I. Alsmadi, and J. Tang, "An Ensemble Feature Selection Method for IoT IDS," in *2020 IEEE 6th International Conference on Dependability in Sensor, Cloud and Big Data Systems and Application (DependSys)*, Dec. 2020, pp. 41–48. doi: 10.1109/DependSys51298.2020.00015.

[13] "NSL-KDD | Datasets | Research | Canadian Institute for Cybersecurity | UNB." https://www.unb.ca/cic/datasets/nsl.html (accessed Nov. 20, 2019).

[14] N. Moustafa and J. Slay, "UNSW-NB15: a comprehensive data set for network intrusion detection systems (UNSW-NB15 network data set)," in *2015 Military Communications and Information Systems Conference (MilCIS)*, Nov. 2015, pp. 1–6. doi: 10.1109/MilCIS.2015.7348942.

[15] A. Alhowaide, I. Alsmadi, and J. Tang, "Features Quality Impact on Cyber Physical Security Systems," in *2019 IEEE 10th Annual Information Technology, Electronics and Mobile Communication Conference (IEMCON)*, Vancouver, BC, Canada, Oct. 2019, pp. 0332–0339. doi: 10.1109/IEMCON.2019.8936280.

[16] "The BoT-IoT Dataset." https://www.unsw.adfa.edu.au/unsw-canberra-cyber/cybersecurity/ADFA-NB15-Datasets/bot_iot.php (accessed Dec. 12, 2019).

[17] A. Alhowaide, "IoT dataset for Intrusion Detection Systems (IDS)." https://kaggle.com/azalhowaide/iot-dataset-for-intrusion-detection-systems-ids (accessed Nov. 09, 2021).